\documentclass[sigconf]{acmart}

\copyrightyear{2026}
\acmYear{2026}
\setcopyright{cc}
\setcctype{by}
\acmConference[GoodIT '26]{International Conference on Information Technology for Social Good}{September 02--04, 2026}{Pisa, Italy}
\acmBooktitle{International Conference on Information Technology for Social Good (GoodIT '26), September 02--04, 2026, Pisa, Italy}
\acmDOI{10.1145/3794786.3830729}
\acmISBN{979-8-4007-2483-1/2026/09}

\begin{document}

\title{Digital Engagement, Income Disparities, and Job Seeking in the United States since 2010}

\author{Shaolong Wu}
\affiliation{%
  \institution{Harvard University}
  \city{Cambridge}
  \state{MA}
  \country{USA}}
\email{lorrywu@g.harvard.edu}

\author{Yijiang River Dong}
\affiliation{%
  \institution{University of Cambridge}
  \city{Cambridge}
  \country{United Kingdom}}
\email{yd358@cam.ac.uk}

\author{Siming He}
\affiliation{%
  \institution{University of California, Berkeley}
  \city{Berkeley}
  \state{CA}
  \country{USA}}
\email{siminghe@berkeley.edu}

\renewcommand{\shortauthors}{Wu, Dong, and He}

\begin{abstract}
Surveys often record how frequently people use the internet without measuring the infrastructures, skills, and support systems that make digital participation possible. Using the U.S. National Longitudinal Survey of Youth 1997 cohort, we study how internet-use frequency relates to labor income, employment attachment, and job seeking after 2010. The main digital-engagement analysis uses the comparable 2011, 2013, and 2015 waves, with 2017 retained as later labor-market context. Across repeated cross sections, daily internet use consistently marks higher income and stronger employment attachment. Relative to daily use, less-than-daily use is associated with roughly 11 to 20 percent lower income, while nonuse is associated with about 18 to 21 percent lower income in 2011 and 2013. Respondents reporting no internet use are also 13 to 23 percentage points less likely to report full-year work. Job-search estimates reveal a distinct mechanism: active search is governed by employment status, search intensity, and application support, so a frequency item sorts respondents more sharply on durable labor-market attachment than on short-window search. Education accounts for a substantial share of the raw digital gradient, and pooled lagged-outcome and doubly robust transition estimates separate durable stratification from positive adoption margins. The results establish internet-use frequency as an informative behavioral marker of digitally mediated labor-market stratification and clarify why routine use should not be treated as a simple measure of digital access.
\end{abstract}



\maketitle

\section{Introduction}
Job search, hiring, scheduling, training, workplace communication, and public-facing employment services increasingly run through digital systems. Online job boards, employer portals, platform labor markets, remote coordination tools, and algorithmic recommendation systems now shape how workers encounter openings and how firms encounter workers \citep{Horton2017,KatzKrueger2019,BarreroBloomDavis2021}. The recruitment pipeline has become even more digitally intensive in recent years, with automated ranking, asynchronous interviews, and AI-assisted screening adding new friction points and new forms of exclusion \citep{FabrisEtAl2025,WuBlumeYeung2026}. That shift makes digital inequality central to labor-market research and creates a recurring measurement problem. Many social surveys observe how often respondents use the internet without observing the infrastructures, affordability conditions, device quality, digital skill, or community support that make that use possible. Broadband coverage, platform literacy, and online behavior are related yet analytically distinct.

This distinction motivates the paper's central question: \emph{what can internet-use frequency reveal about unequal economic opportunity when richer digital-access measures are unavailable?} We study the U.S. National Longitudinal Survey of Youth 1997 cohort (NLSY97), which provides repeated information on income, household conditions, education, employment, job search, and internet-use frequency. We interpret internet-use frequency as \emph{digital engagement}: an observable trace of how routinely respondents participate in institutions that are increasingly organized through digital channels.

In digital-inequality research, the closest parallel is digital capital: the capacity to convert devices, skills, routines, and support into offline advantage \citep{Ragnedda2018}. A frequency measure captures a narrow behavioral slice of digital capital and reveals whether routine online participation is systematically bundled with other forms of social and economic advantage.

The analysis speaks to four related literatures. Labor economics has long shown that education, family structure, and household conditions are strongly related to earnings and employment outcomes \citep{BlauKahn2007,Mincer1996}. That perspective complements work on task-biased technological change and economic identity, which highlights how institutional participation mediates the economic value of technology and skill \citep{AutorLevyMurnane2003,AkerlofKranton2000}. Job-search research emphasizes financial need, search intensity, and self-regulation rather than mere exposure to technology \citep{WanbergWattRumsey1996,CreedEtAl2009}. Employment is a core livelihood infrastructure. Yet technology-led interventions can mistakenly assume that moving services online is equivalent to expanding access. HCI research on employment support shows a more complicated process. Low-resourced job seekers may rely on libraries, employment centers, mentors, and community organizations to find openings, understand application norms, and troubleshoot opaque online systems \citep{DillahuntEtAl2021Survey,DillahuntEtAl2021Centers,IsraniEtAl2021}. Digital exclusion can also create new costs when workers lack privacy, stable devices, documentation, or platform-specific knowledge \citep{DeMarcoEtAl2023}. More generally, observable economic behavior is often socially situated: identity, public evaluation, and institutional context shape how actions translate into opportunity and recognition \citep{AkerlofKranton2000,Wu2025}. Finally, research on broadband expansion shows that connectivity can matter for employment and resilience when infrastructure changes can be identified more credibly \citep{Atasoy2013,Dettling2017}. Our contribution is to show how a widely observed behavioral measure reveals a durable pattern of digitally mediated labor-market stratification when direct measures of infrastructure, skill, and support are unavailable.

We focus on three research questions. How does digital engagement fit into the cross-sectional structure of income, employment attachment, and job search in the 2011--2017 NLSY97 waves? How does that gradient behave when the analysis turns to within-person change and adoption timing? And how should a behavioral internet-use measure be interpreted in social-good research when the available geography is coarse and richer access measures are absent?

The contribution is threefold. First, we show that digital engagement is sharply stratified by education and race/ethnicity before it ever enters a regression model. Second, we show that daily internet use is much more strongly associated with labor income and full-year employment than with active job search. Third, sequential adjustment, lagged-outcome models, fixed-effects extensions, and adoption-timing estimates show that the digital gradient reflects both durable social stratification and positive transition gains from entering daily use. We contribute to the literature by showing internet-use frequency is a behavioral marker of digitally mediated labor-market stratification instead of a mere access variable.

Workforce agencies, public libraries, community colleges, and digital navigator programs often need simple indicators to identify where digital disadvantage is clustered. A low-frequency internet-use measure helps flag bundled disadvantage and directs the next diagnostic question: which input is missing? The binding constraint may be broadband affordability, device reliability, platform literacy, documentation, confidence, private space, or access to a human intermediary who can help translate online access into durable work outcomes. The estimates reported here provide both a clear and substantive baseline from the pre-2020 labor market and a practical interpretation rule for a common survey measure.

\section{Data and analytic design}
We use public-use data from the U.S. National Longitudinal Survey of Youth 1997 cohort, administered by the U.S. Bureau of Labor Statistics. The cohort contains 8,984 respondents born between 1980 and 1984 and first interviewed in 1997. The main analysis focuses on the 2011, 2013, and 2015 waves because these are the waves in which internet-use frequency is measured on a comparable basis. These waves also capture a consequential period in which employment and job search were already digitally mediated during the years before remote work normalized and AI-enabled hiring tools diffused more widely. The 2017 wave is retained as a later labor-market context because the comparable internet-use-frequency item is unavailable in that wave.

The three main outcomes are logged labor income, an indicator for whether the respondent searched for a job in the prior three months, and an employment-attachment indicator for whether the respondent worked at least 50 weeks in the prior year. The employment-attachment outcome matters because it is closer than point-in-time search to livelihood stability and sustained labor-market integration. The focal explanatory variable is internet-use frequency; the same seven-category item appears in the three core waves and is collapsed into daily use, less than daily use, and no use, while no comparable 2017 digital-use category is imputed. Education is collapsed into no high school or GED, high school or associate degree, and bachelor degree or above. We also use broad region, marital status, sex, weeks unemployed in the previous year, and a small set of background covariates. The descriptive analysis additionally reports race and ethnicity categories available in the cohort extracts.

Coverage differs across outcomes. Internet-use frequency is observed for roughly 77--80\% of respondents in the three core waves, labor income for about 51--67\%, and job-search activity for about 54--57\%. Estimation is complete-case within each model. The resulting analytic samples are somewhat more educated, more often married, and more digitally engaged than the full observed waves. The main coefficients therefore describe stratification within a comparatively attached segment of the cohort.

For each wave $t$, the core income specification is
\[
\log(Y_{it}) = \alpha_t + \beta_t D_{it} + \gamma_t X_{it} + \varepsilon_{it},
\]
where $Y_{it}$ is labor income, $D_{it}$ is the internet-use category, and $X_{it}$ contains the structural covariates. For job search and full-year employment, we estimate corresponding logit models and report average marginal effects in percentage points. The repeated-cross-section models ask how digital engagement fits into the observed structure of labor-market inequality once education, marriage, region, and other covariates are held constant.

We then add four extensions that sharpen temporal interpretation. First, we estimate sequentially adjusted income models on a common sample to identify the compositional share of the raw digital gradient. Second, we estimate an individual fixed-effects extension for logged household income and job search over 2011--2017. This is a related within-person test rather than a one-for-one replication of the labor-income models, because household income is more consistently observed in the panel structure available for this extension. Third, we pool the 2011--2013 and 2013--2015 transitions and estimate lagged-outcome models for next-wave labor income, next-wave full-year employment, and next-wave job search. Fourth, we use augmented inverse-probability weighting (AIPW) to compare respondents who move into daily use with observationally similar respondents who remain non-daily. These extensions introduce temporal ordering and identify where a common behavioral survey item is most informative.

\section{Descriptive stratification of digital engagement}
\begin{figure*}[t]
  \centering
  \includegraphics[width=0.96\textwidth]{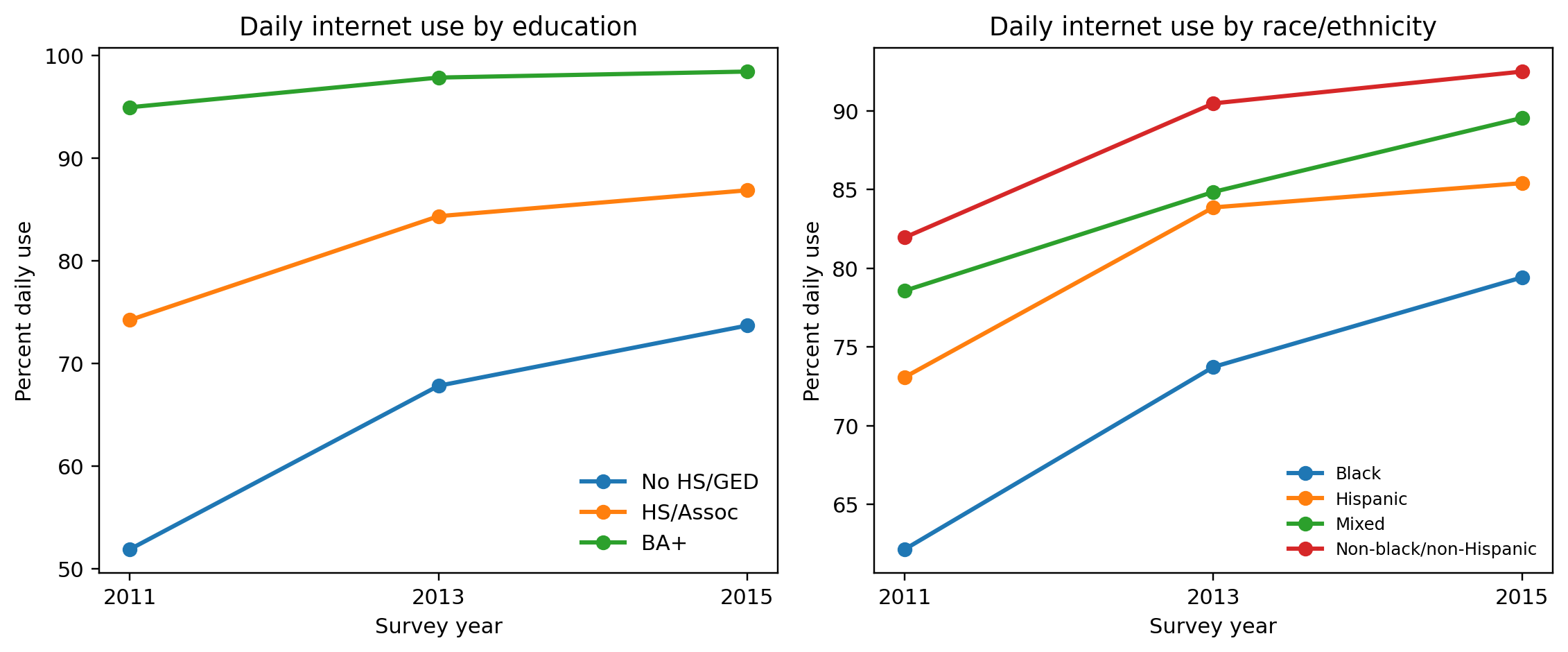}
  \Description{Two line charts. The left chart shows that daily internet use is much more common among respondents with more education in 2011, 2013, and 2015. The right chart shows that daily internet use is also higher among non-Black non-Hispanic and mixed respondents than among Black and Hispanic respondents in all three waves.}
  \caption{Daily internet use is sharply stratified by education and race/ethnicity in the NLSY97 core waves.}
  \label{fig:stratification}
\end{figure*}

Figure~\ref{fig:stratification} shows descriptively that digital engagement is socially located. Daily internet use rises sharply with education. In 2011, the daily-use share is about 52\% among respondents without a high school degree or GED, 74\% among those in the high-school-or-associate category, and 95\% among those with a bachelor degree or above. Race and ethnicity differences are also large. In the same wave, the daily-use share is about 62\% among Black respondents, 73\% among Hispanic respondents, 79\% among mixed-race respondents, and 82\% among non-Black/non-Hispanic respondents. By 2015, daily use rises for every group, and the gaps remain substantial.

Broad place differences are secondary to education and race gradients. The daily-use share ranges from 70.5\% in the South to 78.5\% in the Northeast in 2011, and from 84.2\% in the South to 91.2\% in the West in 2015. Those differences show that place-linked inequality remains part of the story, with the strongest descriptive stratification appearing along education and race/ethnicity.

\begin{figure}[t]
  \centering
  \includegraphics[width=\columnwidth]{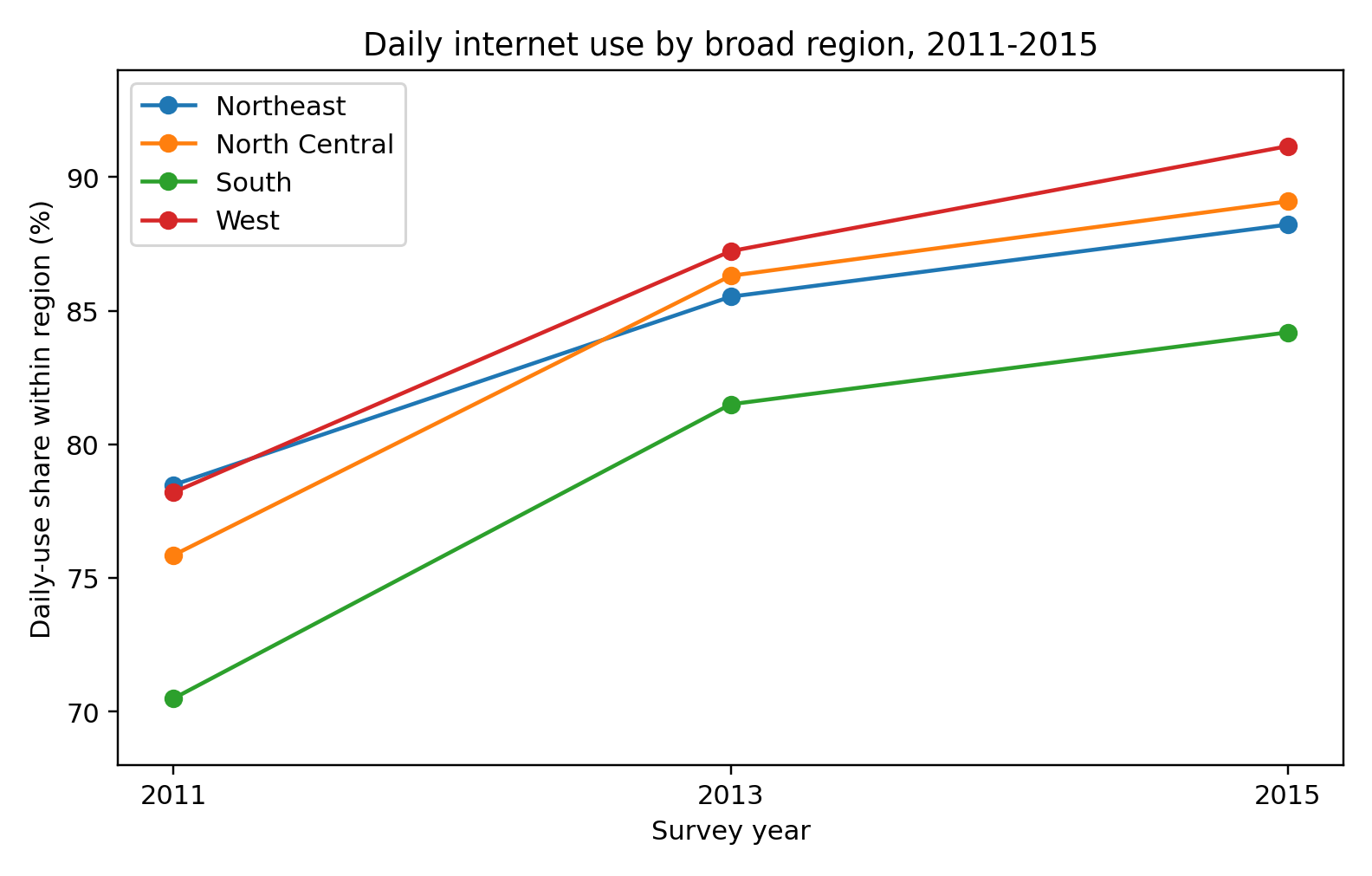}
  \Description{A line chart showing daily internet-use shares for the Northeast, North Central, South, and West in 2011, 2013, and 2015. The South is lowest in every wave and the West is highest by 2015, but the gaps are modest compared with education and race gradients.}
  \caption{Regional differences in daily use are secondary to the education and race gradients.}
  \label{fig:region}
\end{figure}

That contrast is critical: the behavioral internet variable combines digital and social position; geography explains only one part of its structure. It already reflects where respondents sit inside an unequal opportunity structure before it is introduced into any model. The descriptive fact sets up the regression evidence that adjusted digital gradients operate on top of pronounced social stratification.

\section{Results}
\subsection{Cross-sectional digital gradients across outcomes}
\begin{figure*}[t]
  \centering
  \includegraphics[width=0.98\textwidth]{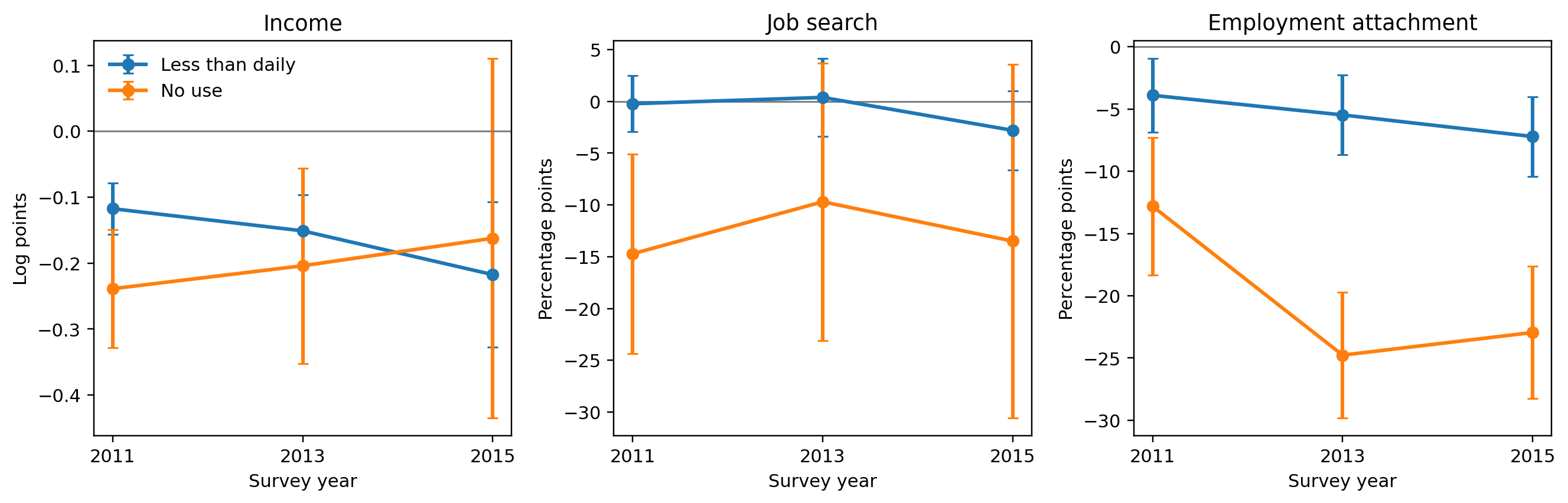}
  \Description{A three-panel figure with error bars by year. The left panel shows negative coefficients for less-than-daily internet use and no internet use on log labor income, relative to daily use. The middle panel shows weak and mostly imprecise effects on job search, except a negative no-use estimate in 2011. The right panel shows stronger negative effects on full-year employment, especially for no use.}
  \caption{Relative to daily use, lower digital engagement is most strongly associated with lower income and lower full-year employment.}
  \label{fig:outcomes}
\end{figure*}

Figure~\ref{fig:outcomes} summarizes the core estimates across the three labor-market outcomes. Education is the strongest positive predictor of labor income and full-year employment in every wave, and marriage is also positively associated with income. The digital-engagement coefficients are meaningful and concentrated in the outcomes that capture durable economic position.

\begin{table*}[t]
\centering
\small
\caption{Main digital-engagement estimates across outcomes}
\label{tab:mainest}
\begin{tabular}{p{2.7cm}p{2.1cm}ccc}
\toprule
Outcome & Contrast relative to daily use & 2011 & 2013 & 2015 \\
\midrule
Labor income & Less than daily & $-0.117$ & $-0.151$ & $-0.217$ \\
Labor income & No use & $-0.239$ & $-0.204$ & $-0.162$ \\
Job search & Less than daily & Near zero & Near zero & Near zero \\
Job search & No use & $-14.7$ pp & Near zero & Near zero \\
Full-year employment & Less than daily & $-3.9$ pp & $-5.5$ pp & $-7.2$ pp \\
Full-year employment & No use & $-12.8$ pp & $-24.8$ pp & $-22.9$ pp \\
\bottomrule
\end{tabular}
\end{table*}

For labor income, the pattern is stable and substantively large. Relative to daily users, less-than-daily users have coefficients of $-0.117$ in 2011, $-0.151$ in 2013, and $-0.217$ in 2015, corresponding to roughly 11\%, 14\%, and 20\% lower income. Respondents reporting no use have coefficients of $-0.239$ in 2011 and $-0.204$ in 2013, or about 21\% and 18\% lower income than otherwise similar daily users. The 2015 no-use coefficient remains negative as nonuse becomes rare by that wave.

\begin{figure}[t]
  \centering
  \includegraphics[width=\columnwidth]{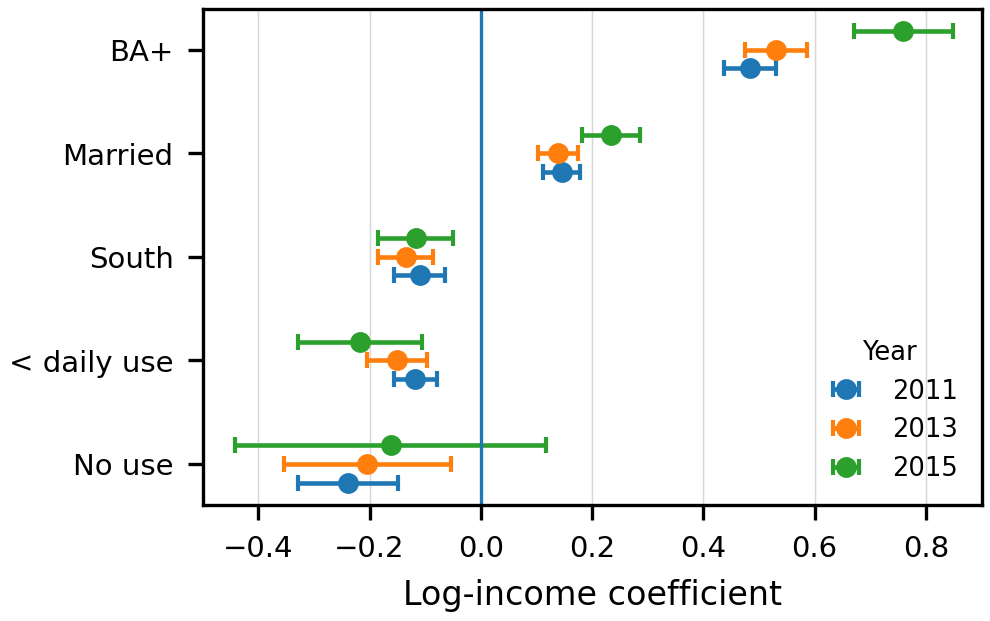}
  \Description{A coefficient plot for the income models showing selected coefficients by year. BA+ is strongly positive in all years, marriage is positive, South is modestly negative, and both less-than-daily use and no use are negative relative to daily use.}
  \caption{Selected coefficients from the income models place the digital gradient alongside durable structural predictors.}
  \label{fig:incomecoef}
\end{figure}

Figure~\ref{fig:incomecoef} places the digital coefficients next to a small set of familiar structural contrasts. Relative to the no-high-school-or-GED category, the bachelor-or-above coefficient is 0.484 in 2011, 0.531 in 2013, and 0.759 in 2015, corresponding to roughly 62\%, 70\%, and 114\% higher labor income. Marriage is also consistently positive, at about 16--26\% higher income, while respondents in the South earn about 10--14\% less than otherwise similar respondents in the Northeast. Digital engagement remains substantively large even when set beside well-known structural divides.

The job-search models identify a different channel. Nonuse is strongly negative in 2011, with an estimated marginal effect of about $-14.7$ percentage points, while less-than-daily use stays close to the daily-use baseline after controls. The most stable positive predictor in these models is bachelor degree attainment, which is associated with about 7.0 percentage points more job search in 2011, 10.9 points more in 2013, and 7.8 points more in 2015. Marriage is generally associated with less search, while weeks unemployed is positively associated with search in the later waves. This combination shows that observed search is shaped by mobility, employment status, and search intensity rather than by digital exposure alone. Active search depends on mechanisms that a frequency item is not designed to observe: platform familiarity, profile quality, occupational targeting, application help, documentation, and the difference between searching while unemployed and searching while employed. That interpretation aligns with HCI research showing that low-resourced job seekers often depend on trusted intermediaries and context-specific assistance in addition to connection \citep{DillahuntEtAl2021Survey,DillahuntEtAl2021Centers,IsraniEtAl2021,DeMarcoEtAl2023}.

Employment attachment provides the clearest social-good outcome. Relative to daily use, less-than-daily use is associated with a 3.9 percentage-point lower probability of full-year employment in 2011, a 5.5 point lower probability in 2013, and a 7.2 point lower probability in 2015. Nonuse is associated with much larger differences: about $-12.8$ points in 2011, $-24.8$ points in 2013, and $-22.9$ points in 2015. In other words, digital engagement sorts respondents more sharply on sustained labor-market integration than on observed search behavior.

\begin{figure}[t]
  \centering
  \includegraphics[width=\columnwidth]{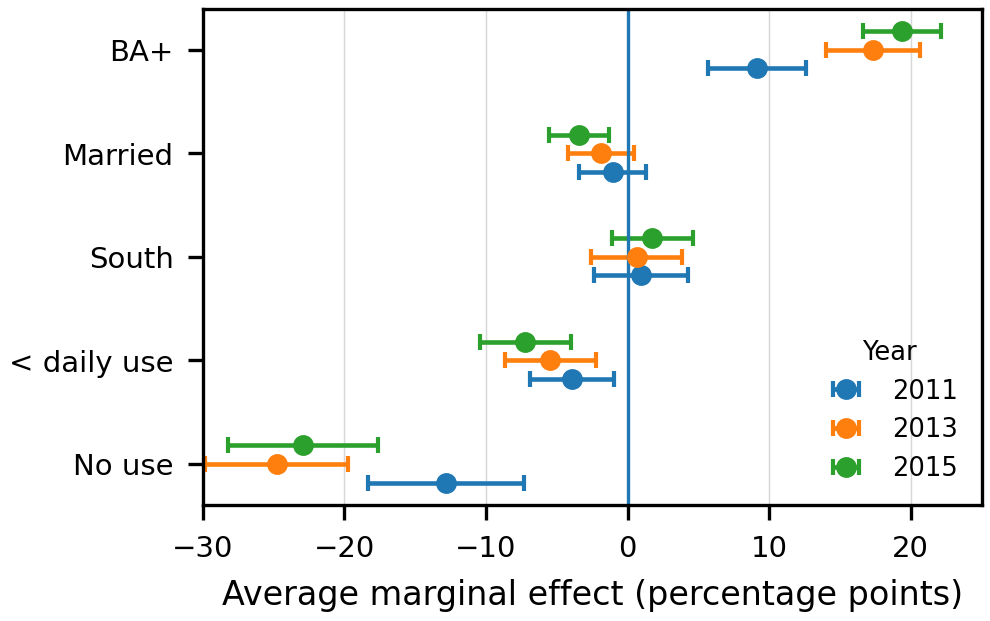}
  \Description{A coefficient plot with selected average marginal effects from the employment-attachment models. Less-than-daily use is modestly negative in every year, no use is strongly negative in every year, and BA plus is strongly positive.}
  \caption{A focused view of the employment-attachment estimates.}
  \label{fig:employmentfocus}
\end{figure}

Figure~\ref{fig:employmentfocus} makes that stability gradient especially easy to read. The less-than-daily penalty is persistent and increasingly negative across the three waves, while the no-use penalty is large throughout. The same figure also shows that BA+ is strongly positive in every year, underscoring that digital engagement and schooling are layered onto one another. This is one reason employment attachment is so informative for social good: it captures whether workers remain connected to work over the year, beyond a single search episode.

Taken together, the cross-sectional results support a disciplined interpretation. Daily internet use is a robust marker of advantage in income and work attachment. The same behavioral item sorts durable labor-market position more sharply than active search once the surrounding opportunity structure is taken into account. Sensitivity checks with explicit race adjustment leave the core income and employment-attachment gradients substantively intact, which identifies compounded rather than redundant disadvantage.

\subsection{The raw digital gradient reflects durable social structure}
\begin{figure*}[t]
  \centering
  \includegraphics[width=0.96\textwidth]{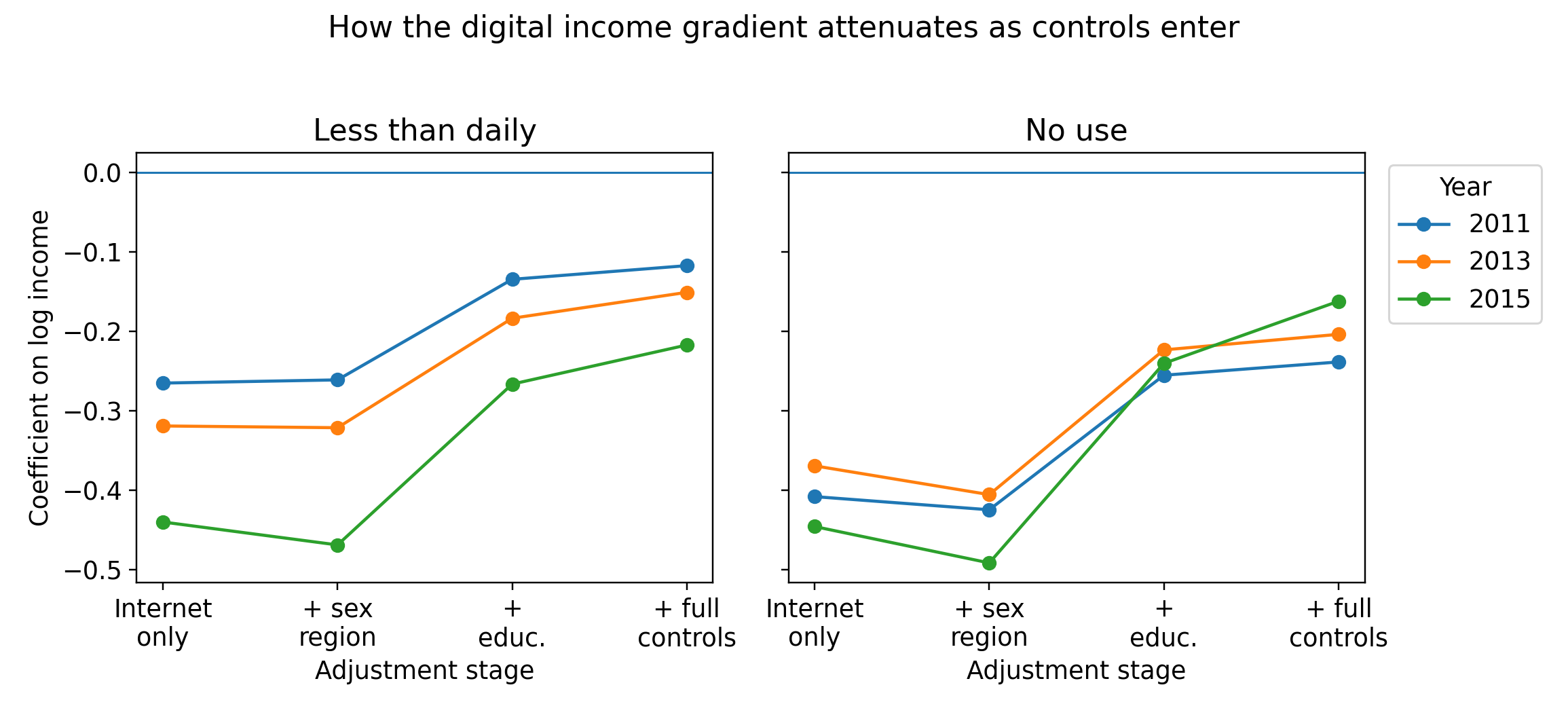}
  \Description{Two line charts, one for less-than-daily use and one for no use, each showing income coefficients by year as controls enter sequentially. The negative coefficients become much smaller after sex, region, education, and full controls are added.}
  \caption{The digital income gradient attenuates substantially as structural controls enter, especially when education is added.}
  \label{fig:attenuation}
\end{figure*}

If internet-use frequency partly bundles durable socioeconomic advantage, the digital coefficients should shrink as structural covariates enter. Figure~\ref{fig:attenuation} shows exactly that pattern for the income models. In 2011, the less-than-daily coefficient moves from $-0.265$ in the unadjusted specification to $-0.117$ in the full model, while the no-use coefficient moves from $-0.408$ to $-0.239$. In 2013, the corresponding changes are from $-0.319$ to $-0.151$ and from $-0.370$ to $-0.204$. In 2015, they are from $-0.440$ to $-0.217$ and from $-0.446$ to $-0.162$.

The largest single attenuation occurs when education enters. We find that internet-use frequency captures durable opportunity structure reflected in schooling and related covariates. The measure still adds independent information after the full covariate set, raising adjusted $R^2$ in the income models by about 0.003 to 0.008 and pseudo $R^2$ in the job-search models by about 0.001 to 0.003. The variable carries substantially more information about income stratification than about active search behavior.

\subsection{Stronger tests of temporal ordering and adoption}
The within-person and prospective extensions separate durable cross-sectional position from shorter-run movement. In the 2011--2017 household-resource fixed-effects extension, changing internet-use frequency is not the main source of movement in logged household income or job-search activity. That contrast is substantively important: the cross-sectional gradient reflects durable differences across respondents, with digital engagement marking position in a broader opportunity structure. Within the same fixed-effects framework, changes in marriage and household composition carry clearer within-person associations with household resources than changes in internet-use frequency do.

The pooled lagged-outcome models reinforce that interpretation. Relative to baseline daily use, baseline less-than-daily use predicts about 0.098 log points lower next-wave labor income and a 7.5 percentage-point lower probability of next-wave full-year employment. Baseline nonuse predicts about 0.103 log points lower next-wave labor income and a 13.5 point lower probability of next-wave full-year employment. The pooled job-search estimates again follow the distinct search mechanism described above. Conditioning on prior outcomes preserves the digital gradient in income and employment attachment. As an additional robustness check, restricting the pooled lagged job-search model to respondents who were not fully employed at baseline leaves the less-than-daily estimate near zero, while baseline nonuse predicts a 7.2 percentage-point lower probability of next-wave job search. The job-search pattern is therefore not an artifact of fully employed respondents having little reason to search. It reflects the distinction between routine digital engagement and the more situational mechanisms that govern active search.

\begin{figure*}[t]
  \centering
  \includegraphics[width=0.96\textwidth]{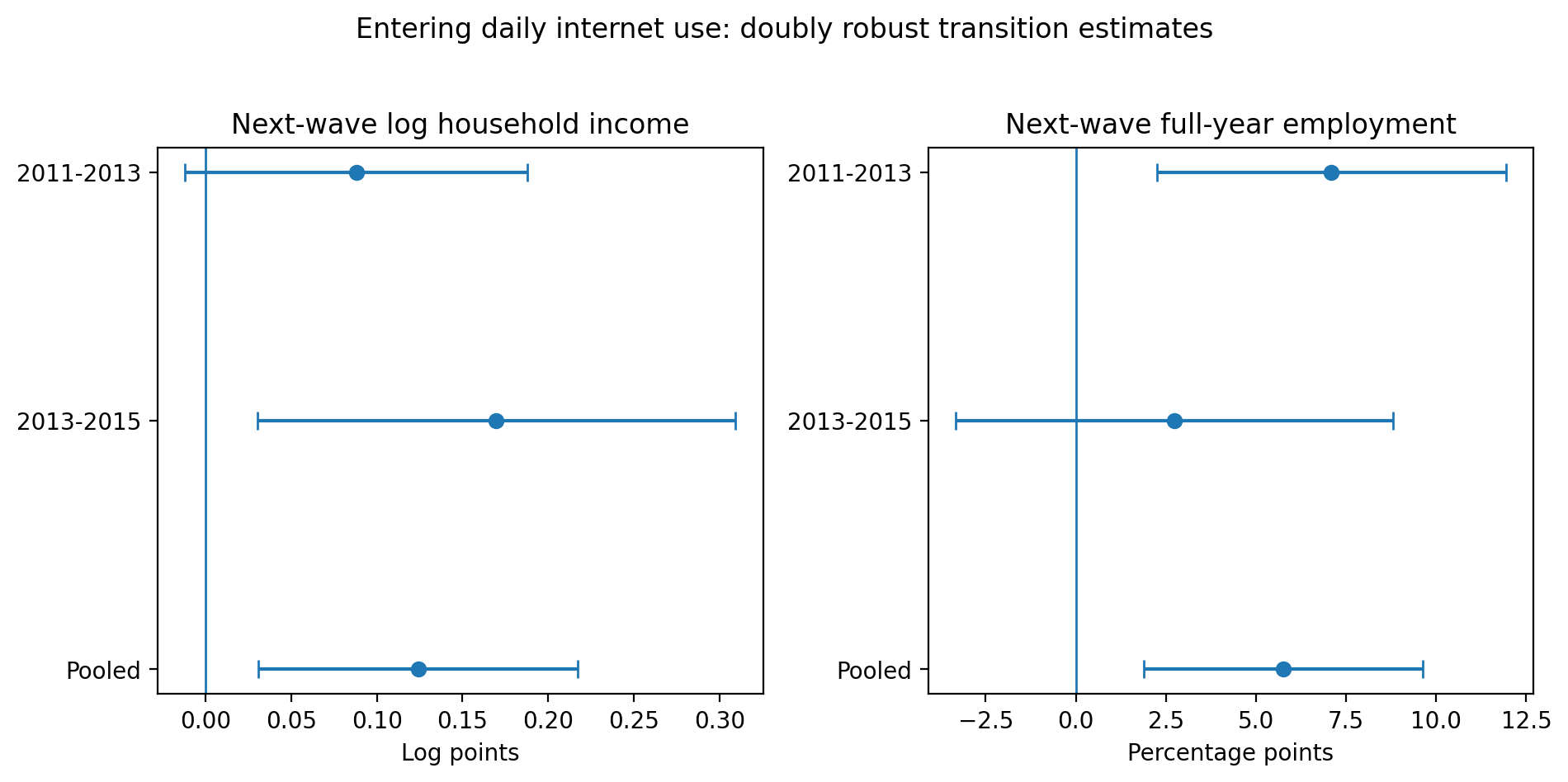}
  \Description{Two panel figure with point estimates and confidence intervals for entering daily internet use. The left panel shows positive effects on next-wave log household income for the pooled sample and for both transition periods. The right panel shows positive effects on next-wave full-year employment, with a larger estimate for 2011 to 2013 than for 2013 to 2015.}
  \caption{Transition into daily internet use predicts positive next-wave income and employment effects.}
  \label{fig:transition}
\end{figure*}

The adoption-timing estimates provide helpful transition-based evidence. Figure~\ref{fig:transition} shows AIPW estimates for respondents who move into daily use relative to comparable respondents who remain non-daily. Entering daily use is associated with about 5.8 percentage points higher next-wave full-year employment and 0.124 log points higher next-wave household income in the pooled estimates.

\begin{table}[t]
  \caption{Selected estimates from temporal-ordering extensions}
  \label{tab:temporal}
  \small
  \centering
  \begin{tabular}{|p{0.70\columnwidth}|r|}
    \hline
    \textbf{Specification and contrast} & \textbf{Estimate} \\
    \hline
    Lagged model: next-wave log labor income, less-than-daily vs. daily & $-0.098$ \\
    Lagged model: next-wave log labor income, no use vs. daily & $-0.103$ \\
    Lagged model: next-wave full-year employment, less-than-daily vs. daily & $-7.5$ pp \\
    Lagged model: next-wave full-year employment, no use vs. daily & $-13.5$ pp \\
    Restricted lagged model: next-wave job search, no use vs. daily & $-7.2$ pp \\
    AIPW transition: enter daily use $\rightarrow$ next-wave log household income & $+0.124$ \\
    AIPW transition: enter daily use $\rightarrow$ next-wave full-year employment & $+5.8$ pp \\
    \hline
  \end{tabular}

  \vspace{2pt}
  {\footnotesize Notes: Pooled lagged-outcome models combine the 2011--2013 and 2013--2015 transitions and condition on the lagged outcome, baseline covariates, and period indicators. The restricted job-search model is limited to respondents who were not fully employed at baseline. AIPW estimates compare entrants into daily use with observationally similar respondents who remain non-daily.}
\end{table}

These transition estimates quantify a positive adoption margin and support our interpretations: entering daily use is associated with better subsequent outcomes, while the repeated cross sections reveal the larger durable opportunity structure in which adoption occurs. Internet-use frequency therefore carries substantive economic information as a marker of digitally mediated stratification.

Taken together, the four extensions impose increasingly demanding tests. Sequential attenuation asks whether the digital coefficients survive compositional adjustment. The fixed-effects models remove stable person-level differences. The lagged-outcome models require digital engagement to predict the next wave beyond the prior level of the outcome. The AIPW transition estimates then isolate the subset of respondents who actually move into daily use and compare them with observationally similar respondents who remain below that threshold. All four designs point in the same qualitative direction: income and employment attachment retain a digital gradient under stronger comparisons. That is exactly what one would expect if routine digital participation matters economically while still being layered onto more durable forms of advantage.

\subsection{Heterogeneity by sex}
\begin{figure}[t]
  \centering
  \includegraphics[width=\columnwidth]{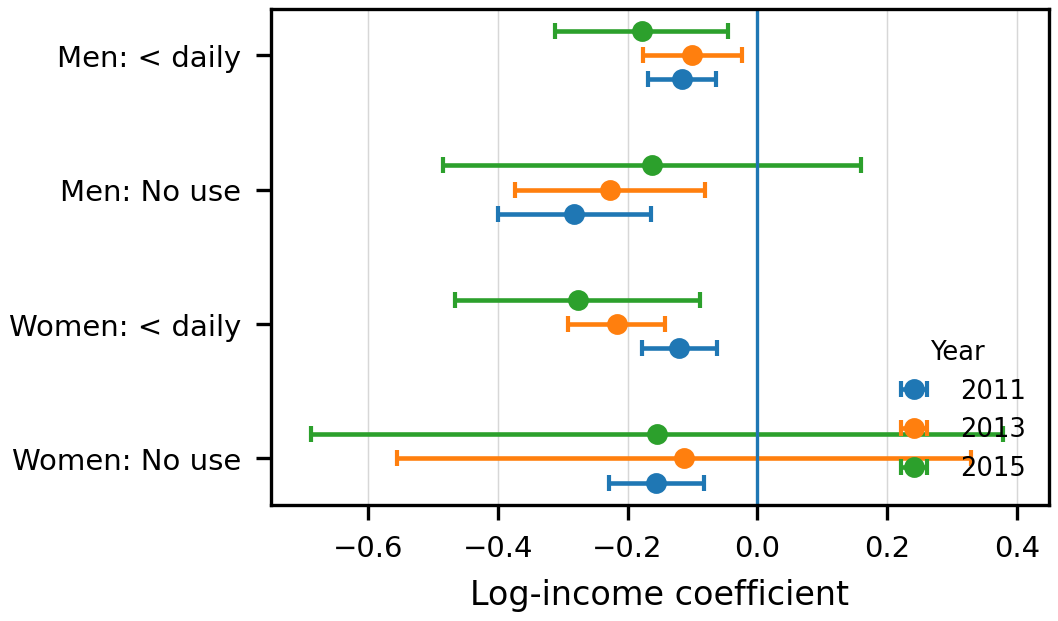}
  \Description{A coefficient plot showing selected income coefficients by sex across waves. For both men and women, less-than-daily use and no use have negative income coefficients relative to daily use, though women have wider confidence intervals.}
  \caption{The digital-engagement income gradient is present for both men and women.}
  \label{fig:sexhetero}
\end{figure}

The income gradient is not confined to one sex. Figure~\ref{fig:sexhetero} plots selected digital-engagement coefficients from sex-stratified income models. For both men and women, the estimated less-than-daily and no-use coefficients are negative across the core waves. The shared pattern shows that routine digital engagement marks a more advantaged earnings position across sex groups, rather than reflecting a single sex-specific labor-market process.

\section{Discussion}
The paper clarifies how a widely used survey measure should be interpreted. Internet-use frequency is a meaningful behavioral marker of digital engagement, distinct from infrastructure access. In the NLSY97 core waves, daily use marks respondents who occupy more advantaged income positions and more stable employment positions in a labor market already structured by digital systems. The job-search results reveal a different mechanism, because active search depends on employment status, search intensity, and application support. The central interpretation is that digital engagement is a bundled marker of position: partly digital, partly social, and embedded in wider structures of schooling, household stability, racialized inequality, and access to help.

The different cross sections establish large contemporaneous differences, the sequential-adjustment models show that education accounts for a large share of those differences, the fixed-effects extension distinguishes durable position from short-run movement, the lagged-outcome models preserve prospective penalties after conditioning on prior outcomes, and the AIPW transition estimates show positive gains from entering daily use. These designs show that routine digital engagement is economically consequential because it runs through wider structures of opportunity.

This cross-design pattern has significant implications for empirical work on digital inequality, where researchers often work with survey and administrative data in the absence of clean infrastructure shocks. The analysis triangulates across repeated cross sections, within-person change, prospective prediction, and doubly robust transition comparisons. These designs point in the same substantive direction: the strongest and most consistent relationships appear for income and employment attachment, while job search follows a distinct mechanism once structural covariates enter. These findings provide a practical template for future research in survey and administrative settings where direct measures of infrastructure, skill, and support remain limited.

Daily internet use is best understood not as a narrow access variable and not as a generic proxy for socioeconomic status, but as an observable trace of digital capital: repeated online routines that reflect the capacity to convert digital participation into labor-market benefit \citep{Ragnedda2018}. The measure is therefore partly digital and partly social. It indexes regularity, confidence, and embeddedness in digitally mediated institutions, but it does so through a bundle that includes schooling, household stability, racialized inequality, and access to help. The stronger association with employment attachment than with job search is especially revealing. Stable work attachment is a stronger livelihood-stability outcome than point-in-time search, and digital engagement sorts respondents much more sharply on that dimension. The job-search pattern shows that active search depends on mechanisms outside the frequency measure: trusted intermediaries, resume assistance, profile management, application coaching, and platform-specific know-how.

This distinction also changes how low internet-use reports should be interpreted in practice. For workforce programs, libraries, community colleges, and digital-inclusion initiatives, low reported use flags a broader bundle of constraints: device quality, unstable service, limited digital skill, low confidence, limited application support, lack of private space, and limited access to trusted intermediaries who help people navigate online systems. We find that internet-use frequency is more strongly associated with employment attachment than with job search. The estimates point to a central channel: stronger digital routines help people sustain connection to work over time.

The race and region results deepen that interpretation. Black respondents had much lower daily-use shares than non-Black/non-Hispanic respondents throughout the core period, yet the digital gradient in income and employment attachment survives explicit race adjustment. Those facts identify compounded disadvantage rather than simple redundancy. Race shapes who is more likely to be digitally engaged in the first place, while digital engagement continues to differentiate labor-market outcomes within racial groups. Broad region works differently. The South consistently shows the lowest daily-use share, and the regional gradient is secondary to education and race. That is precisely what one would expect if broad Census regions capture only part of the place dimension of digital inequality and if the survey variable is picking up a wider opportunity structure than infrastructure alone.

\begin{figure}[t]
  \centering
  \includegraphics[width=\columnwidth]{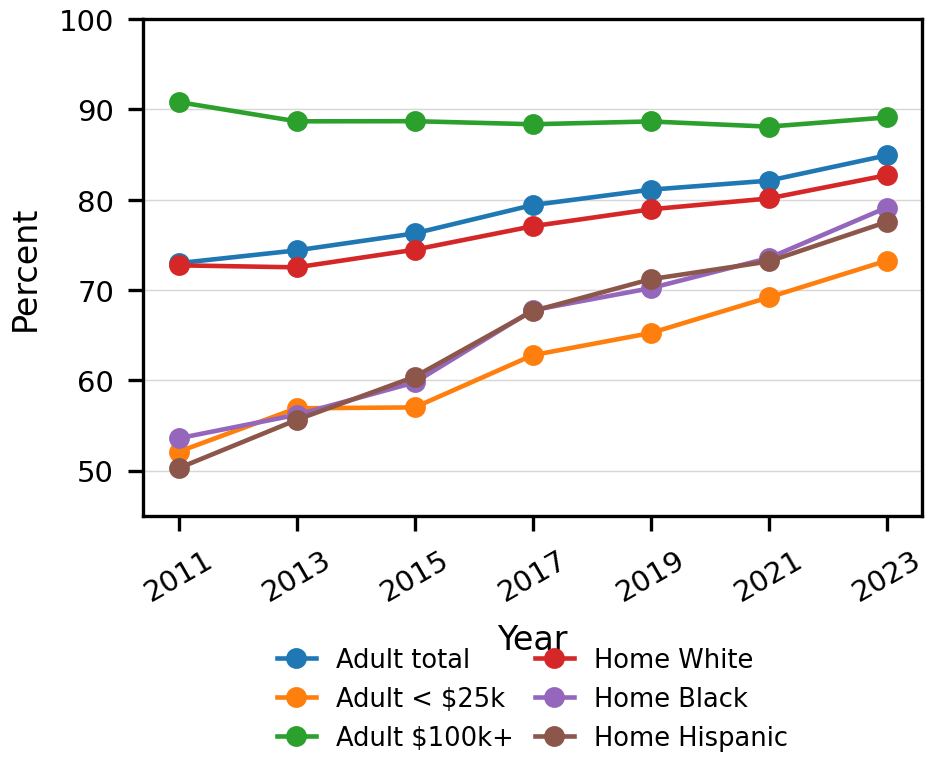}
  \Description{A line chart with six national benchmark series from 2011 to 2023. Overall adult internet use rises over time, but a large gap remains between adults under 25,000 dollars household income and adults above 100,000 dollars. Home internet use also rises for White, Black, and Hispanic adults, while racial gaps persist.}
  \caption{Selected NTIA public benchmark series show broad diffusion of internet use after the study period alongside persistent inequality by income and race.}
  \label{fig:ntia}
\end{figure}

In this paper, these estimates come from 2011--2017, before remote-work normalization and before AI-enabled hiring tools became routine across the recruitment pipeline. Set against later public benchmark series, the present results provide a useful pre-AI baseline for today's labor market. Figure~\ref{fig:ntia} places the cohort results in national context. Adult internet use rises from 73.0\% in 2011 to 84.9\% in 2023, yet adults under \$25,000 remain well below adults above \$100,000 in every benchmark year. Home internet use also rises for White, Black, and Hispanic adults, but substantial gaps remain in 2023 \citep{NTIA2024}. Later studies should test whether AI-assisted search, automated screening, and app-based scheduling strengthen the digital gradient in search behavior more than in this pre-2020 setting. Even if the mechanisms change, the measurement lesson should remain the same: behavioral internet-use variables are most useful when treated as evidence of bundled opportunity structures, not as substitutes for the infrastructures and support ecosystems that policy is trying to build.

The 2017 contextual extension points in the same direction. The comparable digital-engagement item is not available, but the broader labor-market backdrop remains recognizable:
\begin{figure}[t]
  \centering
  \includegraphics[width=\columnwidth]{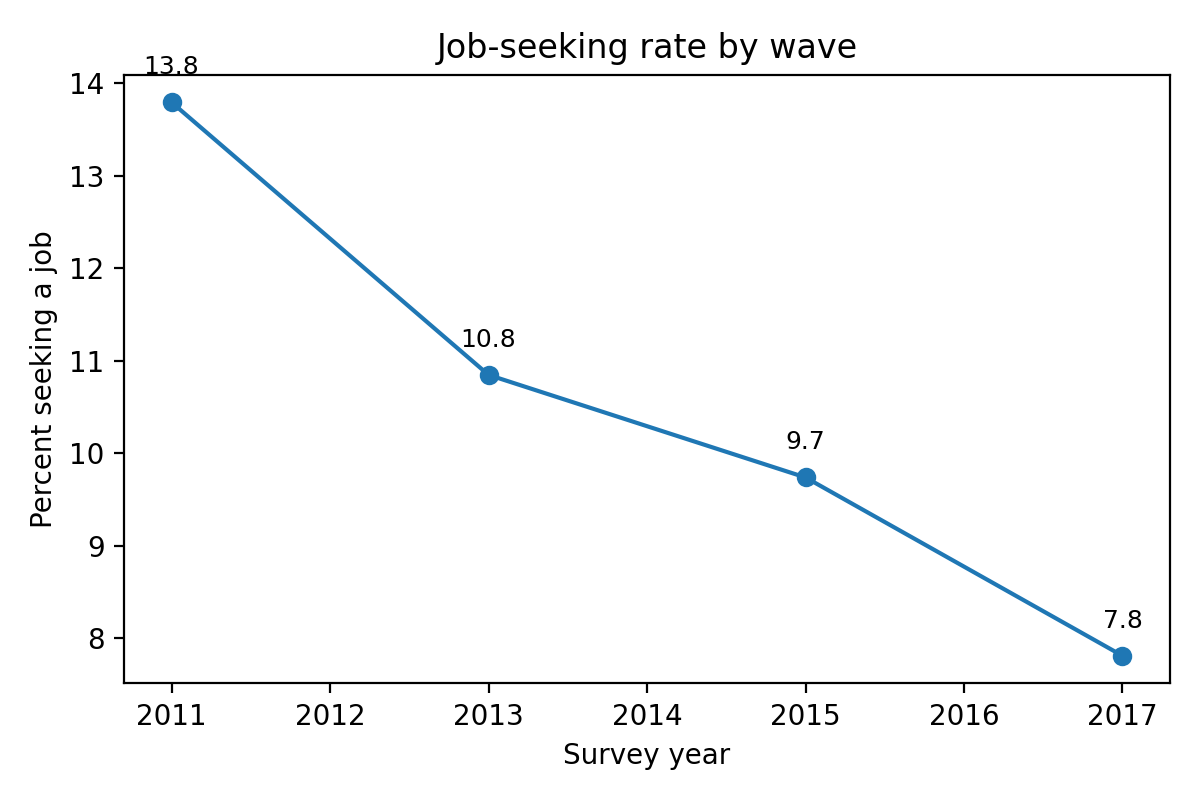}
  \Description{A line chart showing the share of respondents seeking a job in 2011, 2013, 2015, and 2017. The rate declines steadily from 13.8 percent in 2011 to 7.8 percent in 2017.}
  \caption{Job-seeking activity declines across the later wave retained for contextual extension.}
  \label{fig:jobrate}
\end{figure}

Figure~\ref{fig:jobrate} shows that the overall job-seeking rate falls from 13.8\% in 2011 to 10.8\% in 2013, 9.7\% in 2015, and 7.8\% in 2017. The job-search pattern therefore appears in a period when search becomes less common overall even as labor-market mediation through digital systems continues to deepen.

Future work should move toward more community-embedded forms of measurement. Follow-up work should combine neighborhood or metro-level broadband conditions with data from workforce boards, libraries, community colleges, digital navigator programs, and local employers. That design would make it possible to observe who receives help with resumes, online assessments, video interviews, and platform onboarding; where applicants get stuck; and which institutions compensate most effectively for low digital engagement. Although this paper is U.S.-focused, the distinction between engagement and access should travel beyond the U.S. as well, especially in settings where connectivity is intermittent, shared, or institutionally mediated.

\section{Conclusion}
Across the 2011, 2013, and 2015 NLSY97 waves, daily internet use consistently marks a more advantaged labor-market position, especially for labor income and full-year employment. The full evidence shows that this gradient is compositional, durable, and reinforced by positive adoption margins. Internet-use frequency is therefore best understood as digital engagement: an informative behavioral marker of stratified participation in digitally mediated labor markets.

This distinction clarifies how a widely used survey measure should be interpreted in applied social science research on technology access and downstream social implications. Internet-use frequency captures a behavioral dimension of digital engagement; diagnosing the precise binding input requires additional evidence on broadband, devices, skills, platform literacy, and access to trusted intermediaries. That matters for policy as well as measurement. Efforts to expand digital inclusion are likely to be most effective when connectivity is paired with the skills, support, and institutional access needed to translate online participation into durable labor-market opportunity.


\begin{thebibliography}{99}

\bibitem[Akerlof and Kranton(2000)]{AkerlofKranton2000}
George A. Akerlof and Rachel E. Kranton. 2000. Economics and Identity. \textit{Quarterly Journal of Economics} 115(3), 715--753.

\bibitem[Autor et~al.(2003)]{AutorLevyMurnane2003}
David H. Autor, Frank Levy, and Richard J. Murnane. 2003. The Skill Content of Recent Technological Change: An Empirical Exploration. \textit{Quarterly Journal of Economics} 118(4), 1279--1333.

\bibitem[Atasoy(2013)]{Atasoy2013}
Hilal Atasoy. 2013. The Effects of Broadband Internet Expansion on Labor Market Outcomes. \textit{ILR Review} 66(2), 315--345.

\bibitem[Barrero et~al.(2021)]{BarreroBloomDavis2021}
Jose Maria Barrero, Nicholas Bloom, and Steven J. Davis. 2021. Internet Access and its Implications for Productivity, Inequality, and Resilience. \textit{NBER Working Paper} 29102.

\bibitem[Blau and Kahn(2007)]{BlauKahn2007}
Francine D. Blau and Lawrence M. Kahn. 2007. Changes in the Labor Supply Behavior of Married Women: 1980--2000. \textit{Journal of Labor Economics} 25(3), 393--438.

\bibitem[Creed et~al.(2009)]{CreedEtAl2009}
Peter A. Creed, Vivien King, Michelle Hood, and Robert McKenzie. 2009. Goal Orientation, Self Regulation Strategies, and Job Seeking Intensity in Unemployed Adults. \textit{Journal of Applied Psychology} 94(3), 806--813.

\bibitem[De Marco et~al.(2023)]{DeMarcoEtAl2023}
Stefano De Marco, Guillaume Dumont, Ellen Johanna Helsper, Alejandro D\'iaz-Guerra, Mirko Antino, Alfredo Rodr\'iguez-Mu\~noz, and Jos\'e-Luis Mart\'inez-Cantos. 2023. Jobless and Burnt Out: Digital Inequality and Online Access to the Labor Market. \textit{Social Inclusion} 11(4), 184--197.

\bibitem[Dettling(2017)]{Dettling2017}
Lisa J. Dettling. 2017. Broadband in the Labor Market: The Impact of Residential High Speed Internet on Married Women's Labor Force Participation. \textit{ILR Review} 70(2), 451--482.

\bibitem[Dillahunt et~al.(2021a)]{DillahuntEtAl2021Survey}
Tawanna R. Dillahunt, Aarti Israni, Alex Jiahong Lu, Mingzhi Cai, and Joey Chiao-Yin Hsiao. 2021. Examining the Use of Online Platforms for Employment: A Survey of U.S. Job Seekers. In \textit{Proceedings of the 2021 CHI Conference on Human Factors in Computing Systems}, Article 562, 1--23.

\bibitem[Dillahunt et~al.(2021b)]{DillahuntEtAl2021Centers}
Tawanna R. Dillahunt, Matthew Garvin, Marcy Held, and Julie Hui. 2021. Implications for Supporting Marginalized Job Seekers: Lessons from Employment Centers. \textit{Proceedings of the ACM on Human-Computer Interaction} 5(CSCW2), Article 324, 1--24.

\bibitem[Fabris et~al.(2025)]{FabrisEtAl2025}
Alessandro Fabris, Nina Baranowska, Matthew J. Dennis, David Graus, Philipp Hacker, Jorge Saldivar, Frederik J. Zuiderveen Borgesius, and Asia J. Biega. 2025. Fairness and Bias in Algorithmic Hiring: A Multidisciplinary Survey. \textit{ACM Transactions on Intelligent Systems and Technology} 16(1), Article 16, 1--54.

\bibitem[Horton(2017)]{Horton2017}
John J. Horton. 2017. The Effects of Algorithmic Labor Market Recommendations: Evidence from a Field Experiment. \textit{Journal of Labor Economics} 35(2), 345--385.

\bibitem[Israni et~al.(2021)]{IsraniEtAl2021}
Aarti Israni, Nicole B. Ellison, and Tawanna R. Dillahunt. 2021. `A Library of People': Online Resource-Seeking in Low-Income Communities. \textit{Proceedings of the ACM on Human-Computer Interaction} 5(CSCW1), Article 152, 1--28.

\bibitem[Katz and Krueger(2019)]{KatzKrueger2019}
Lawrence F. Katz and Alan B. Krueger. 2019. Understanding Trends in Alternative Work Arrangements in the United States. \textit{RSF: The Russell Sage Foundation Journal of the Social Sciences} 5(5), 132--146.

\bibitem[Mincer(1996)]{Mincer1996}
Jacob Mincer. 1996. Economic Development, Growth of Human Capital, and the Dynamics of the Wage Structure. \textit{Journal of Economic Growth} 1(1), 29--48.

\bibitem[NTIA(2024)]{NTIA2024}
National Telecommunications and Information Administration. 2024. \textit{NTIA Internet Use Survey Analyze Table}. U.S. Department of Commerce.

\bibitem[Ragnedda(2018)]{Ragnedda2018}
Massimo Ragnedda. 2018. Conceptualizing Digital Capital. \textit{Telematics and Informatics} 35(8), 2366--2375.

\bibitem[Wanberg et~al.(1996)]{WanbergWattRumsey1996}
Connie R. Wanberg, John D. Watt, and Deborah J. Rumsey. 1996. Individuals Without Jobs: An Empirical Study of Job Seeking Behavior and Reemployment. \textit{Journal of Applied Psychology} 81(1), 76--87.

\bibitem[Wu(2025)]{Wu2025}
Shaolong Wu. 2025. Are ESG Improvements Recognized? Perspectives from the Public Sentiments. \textit{The Journal of Impact and ESG Investing} 5(4), 24--51.

\bibitem[Wu et~al.(2026)]{WuBlumeYeung2026}
Shaolong Wu, James Blume, and Geshi Yeung. 2026. Alternative Fairness and Accuracy Optimization in Criminal Justice. In \textit{AAAI 2026, AIGOV}. Singapore. arXiv:2511.04505.

\end{thebibliography}
\end{document}